\documentclass[11pt]{article}
\usepackage{moriond,epsfig}

\def\be{\begin{equation}}
\def\ee{\end{equation}}
\def\bea{\begin{eqnarray}}
\def\eea{\end{eqnarray}}

\begin{document}
\vspace*{4cm}
\title{Searches for New Phenomena at the LHC}

\author{Lars Sonnenschein \\
on behalf of the CMS and ATLAS collaborations}

\address{RWTH Aachen University, III. Phys. Inst. A,\\
52056 Aachen, Germany \\
Submitted to conference proceedings of Rencontres de Moriond QCD and High Energy Interactions 2012 \\
(Conference report CMS CR-2012/045)}

\maketitle\abstracts{
Searches for physics beyond the Standard Model (SM) with the CMS~\cite{cms} and ATLAS~\cite{atlas} experiments 
in $pp$ collisions at a centre of mass energy of $\sqrt{s}=7$~TeV at the LHC are presented. 
The discussed results are based on data taken in 2011, making use of integrated luminosities between $\cal{L}=$1.1 and 4.9 fb$^{-1}$.
Various important theories, encompassing 
TeV scale gravity, quark/lepton compositeness, contact interactions, new
heavy vector bosons and other exotic signatures are probed.
}

\section{Introduction}
In the following is focused on non-resonant search channels 
where the invariant mass of a new particle can not be fully reconstructed
due to its decay modes including undetected daughter particles 
\begin{figure}[t]
\vspace*{-7ex}
\hspace*{-23ex}
\psfig{figure=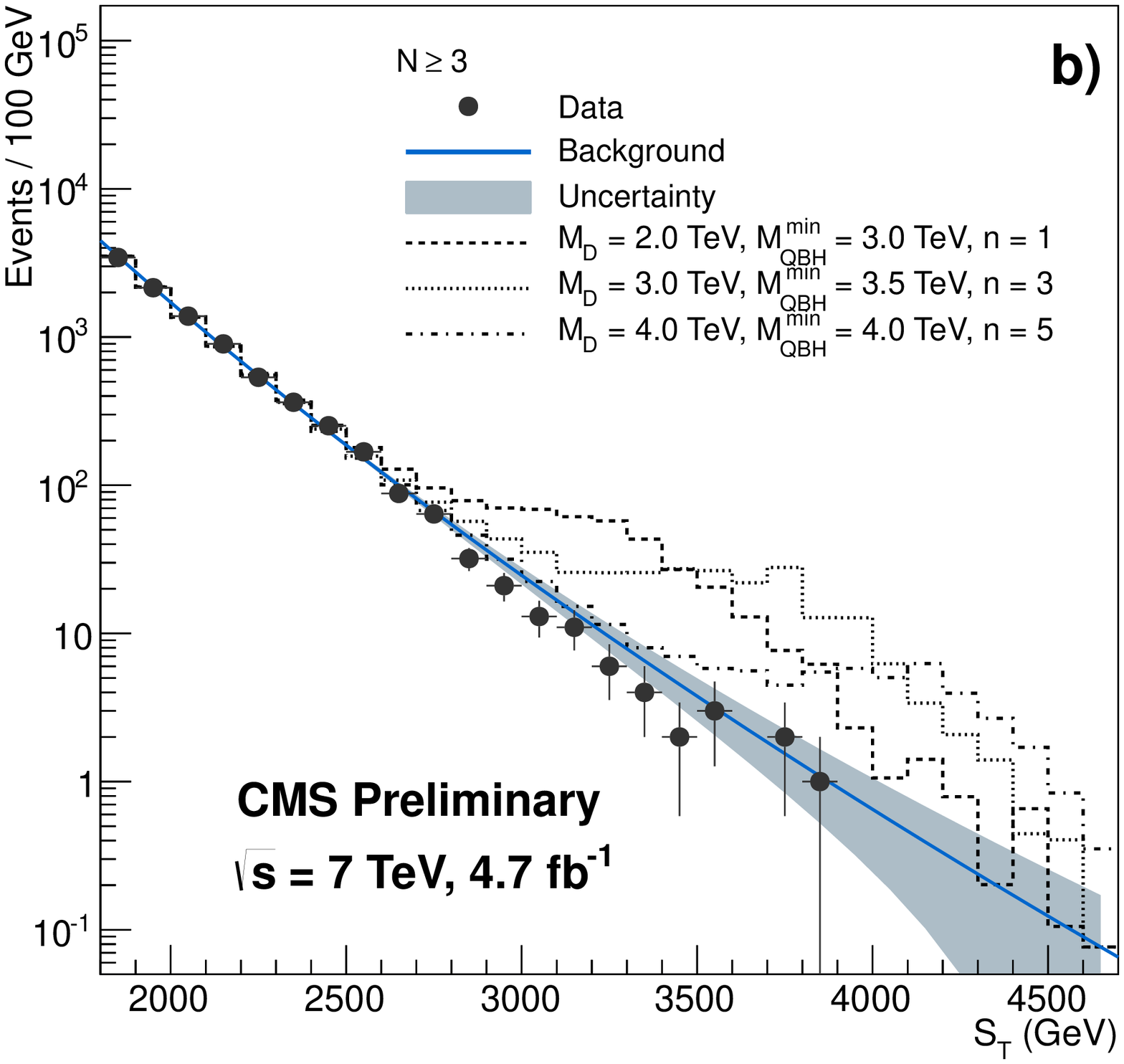,height=4.4in}
\hspace*{-2.0ex}
\psfig{figure=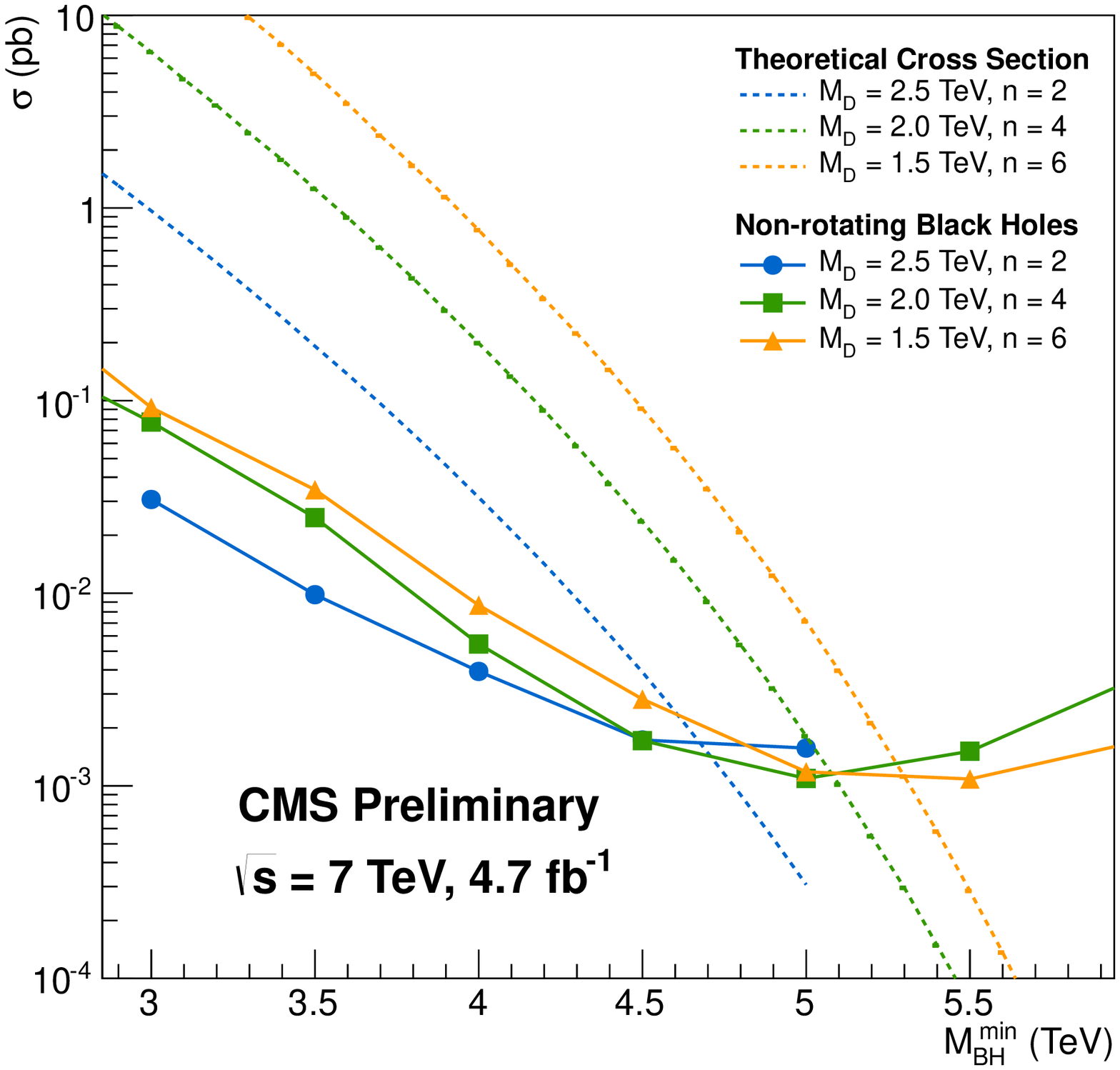,height=4.4in}
\vspace*{-4ex}
\caption{\label{bh}
Left: $S_T$ distribution in the $N\geq3$ final state objects bin for data 
superposed by the background prediction with uncertainties. Some simulated signals with 
multidimensional Planck scale $M_D=2-4$~TeV are indicated, too. At the right are shown cross 
section limits at 95\% C.L. as a function of minimum black-hole mass for various black hole 
parameter sets. 
}
\end{figure}
\begin{figure}[t]
\hspace*{-6ex}
\psfig{figure=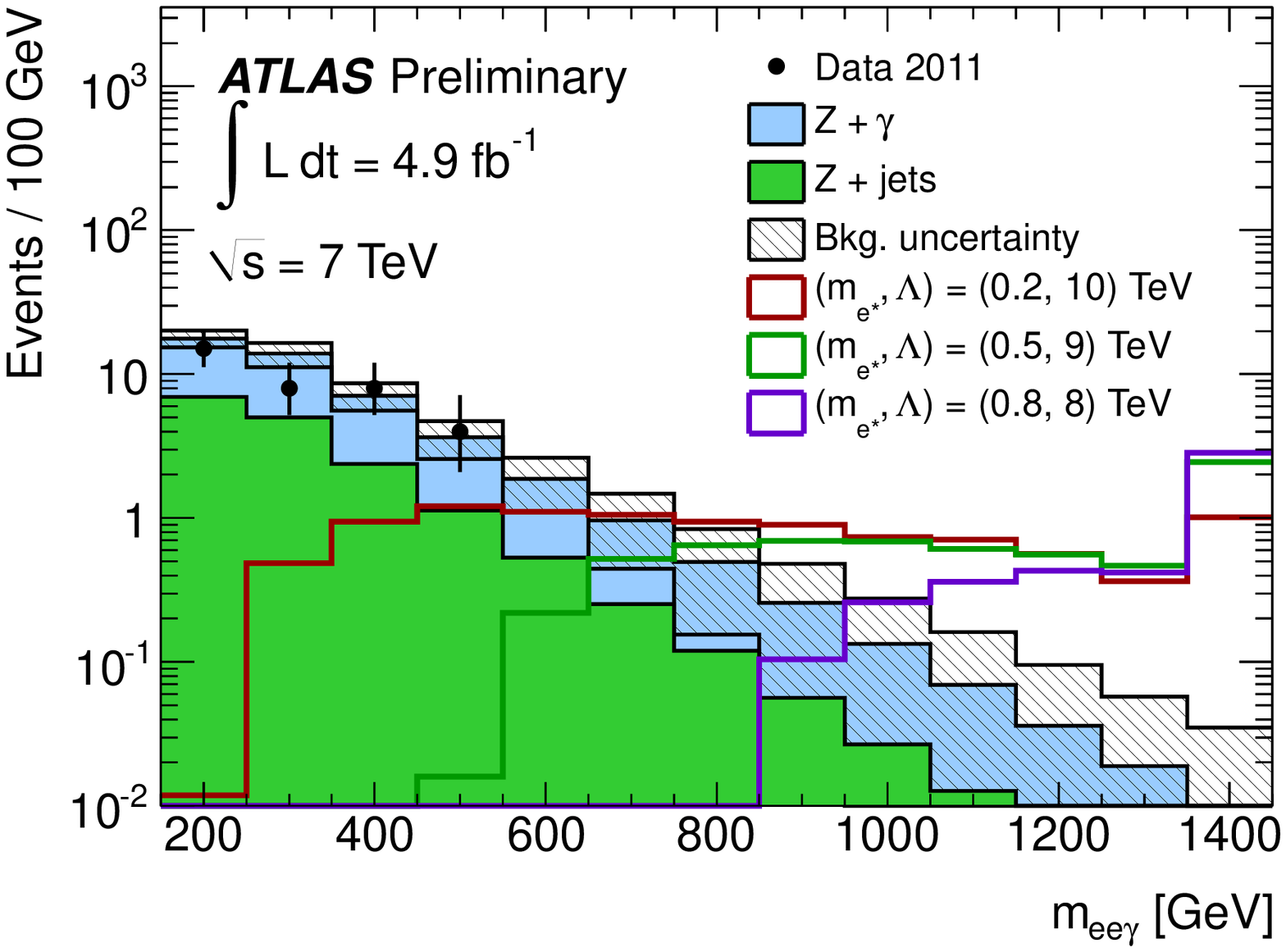,height=2.5in}
\hspace*{-3ex}
\psfig{figure=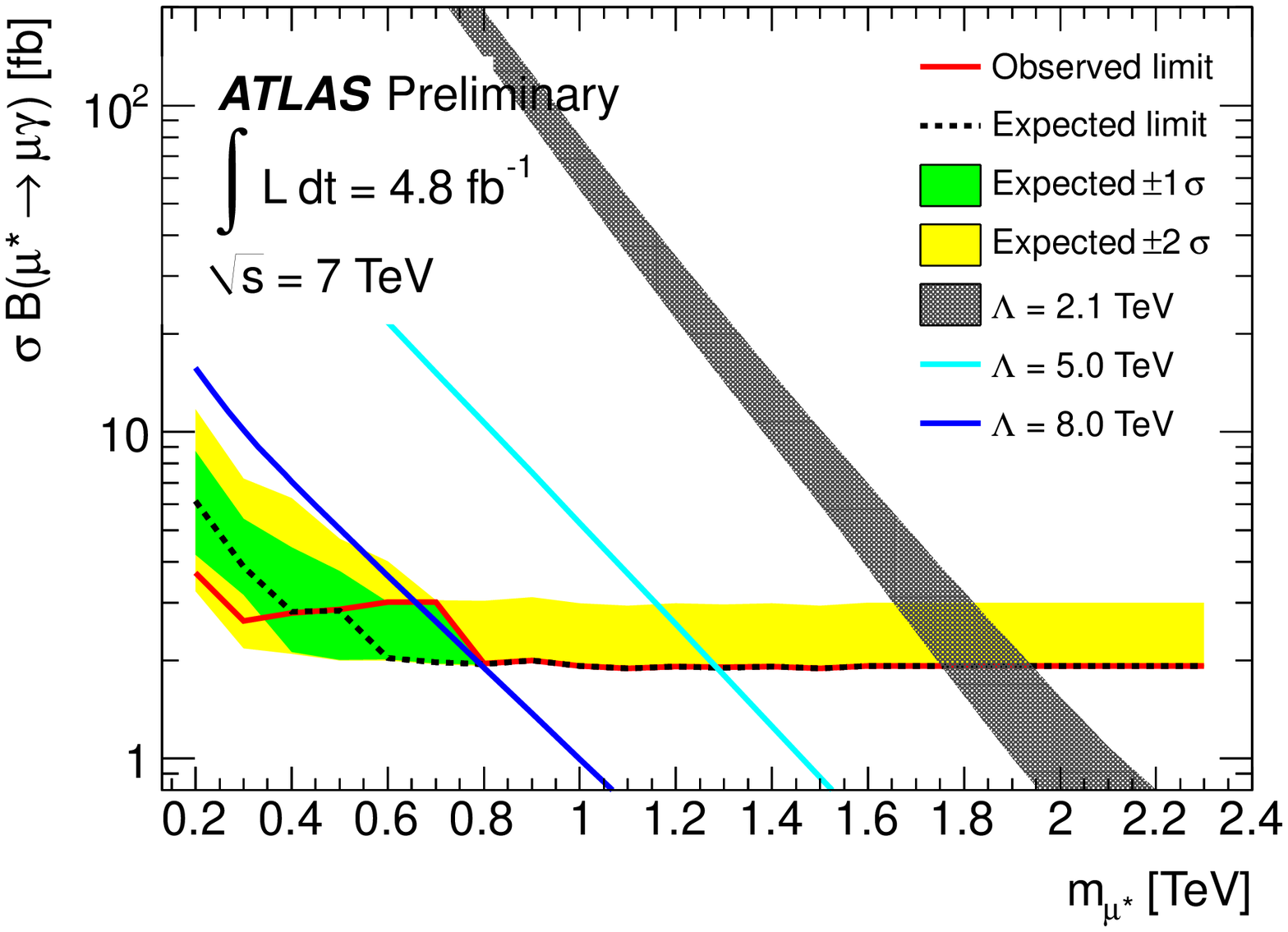,height=2.5in}
\hspace*{8ex}
\vspace*{-4ex}
\caption{\label{emustar}
Left: $ee\gamma$ invariant mass distribution of data in comparison to simulation. 
In the right plot are shown the 95\% C.L. exclusion limits on the excited muon production 
cross section times branching ratio as a function of its mass. Indicated are also three 
different compositeness scales $\Lambda$.
}
\end{figure}
or where the signal does not consist of the production of a resonant particle.
Only new results (after the HCP conference in November 2011) are presented.
A complete list of public analysis results in exotic searches for new physics 
(preliminary, published and submitted or accepted for publication) 
can be found in ref. \cite{CMS_ExoPub} and ref. \cite{ATLAS_ExoPub}.

The next section covers the search for TeV scale gravity. 
The subsequent searches are dedicated to numerous different 
model interpretations and are categorised according to their final states 
into lepton production (section \ref{lepProd}), lepton plus jet production
(section \ref{lepJetProd}) and jet production (section \ref{jetProd}).
In each category only one analysis is discussed here exemplarily. Further presented 
analyses are cited accordingly.
Throughout this article the convention $c \equiv 1$ is adopted for the speed of light.

\section{TeV scale gravity} \label{TeVgravity}

The CMS search~\cite{cms_exo-11-071} for microscopic black holes is based on ${\cal{L}}=4.7$~fb$^{-1}$.
Energetic multiparticle final states including jets, bosons and leptons
are selected by means of the scalar transverse momentum sum $S_T$, taking into
account also the missing transverse energy $E_T^{\mbox{\scriptsize miss}}$ of the 
event. The left plot in Fig.~\ref{bh} shows the $S_T$ distribution of the data for the
$N\geq3$ final state object multiplicity bin, 
together with the predicted background which is dominated by multijet 
production and has been estimated by means of the data.

Limits are set on production cross sections (Fig.~\ref{bh}, right) as a function of the
minimum black-hole mass.
These limits are interpreted in terms of minimal 
Quantum Black Hole masses $m_{\mbox{\scriptsize QBH}}^{\min}$ as a function of the 
multidimensional Planck mass $M_D$ for several extra dimensions.
Further limits on minimum string-ball mass and semi-classical black hole mass
$m_{\mbox{\scriptsize BH}}^{\min}$ are estimated, keeping in mind that the model 
validity breaks down for $m_{\mbox{\scriptsize BH}}^{\min}\simeq 3 - 5 M_D$.

\noindent
Further new results in search of black holes, extra dimensions, dark matter and unparticles
~\cite{atlas-conf-2011-147}~\cite{cms_exo-11-038}~\cite{jhep-5-85-2011}~\cite{atlas_arXiv1112.2194}~\cite{cms-exo-11-087}
~\cite{cms-exo-11-096}~\cite{cms-exo-11-059}~\cite{cms-exo-11-061} have been presented.

\section{Searches in lepton production} \label{lepProd}

The ATLAS search~\cite{atlas-conf-2012-008} for excited leptons $\ell^*\rightarrow\ell\gamma, \ell=e,\mu$ is an update of the previous measurement~\cite{atlas-hepex1201.3293} and makes use of ${\cal{L}}_{ee(\mu\mu)}=4.9(4.8)$~fb$^{-1}$. The excited leptons are expected in the 
electromagnetic radiative decay 
channel $\ell^*\rightarrow \ell\gamma$, produced together with a charge conjugated same flavour lepton
via a four-fermion contact interaction at a given compositeness scale $\Lambda$. The dominant
background consists of Drell-Yan production plus an additional photon or jet. All background predictions
are evaluated with simulated samples. Background from multijets and semileptonic heavy flavour decays
is heavily suppressed by isolation requirements. In Fig.~\ref{emustar} left plot the invariant 
dilepton photon mass distribution is shown for the electron channel.
The signal search region is defined by a sliding lower threshold of $m_{\ell\ell\gamma}>m_{\ell^*}+150$~GeV.
95\% C.L. exclusion limits on the production cross section times branching ratio as a function of 
the excited muon invariant mass are shown in Fig.~\ref{emustar}, right plot. 
For $m_{\ell^*}>0.9$~TeV the observed upper limits on $\sigma\times BR$ are 1.0~fb and 1.9~fb
in the $e^*$ and $\mu^*$ channels, respectively. These limits are translated into bounds on the
compositeness scale $\Lambda$ as a function of the excited lepton mass. For $\Lambda=m_{\ell^*}$
masses below 2.0~TeV and 1.9~TeV are excluded for the $e^*$ and $\mu^*$ channels, respectively. 

\noindent
Further new results in lepton 
production~\cite{cms-exo-11-024}~\cite{cms-exo-11-045}~\cite{atlas-hepex1112.4462}
have been presented.

\section{Searches in lepton + jet production} \label{lepJetProd}

The CMS search~\cite{cms-exo-11-036} for heavy bottom like quarks is based on ${\cal{L}}=4.6$~fb$^{-1}$.
These $b'$ quarks are assumed to decay exclusively to $tW$. Lighter  $b'$ quarks are disfavoured
by results form previous experiments. The pair production
$b'\bar{b'}\rightarrow tW^-\bar{t}W^+$ can be identified by the distinctive signatures of
trileptons or same-sign dileptons, both accompanied by at least one $b$-jet.
Jets are reconstructed with the anti-$k_T$ jet algorithm making use of the distance measure $R=0.5$ in 
rapidity $y$, azimuthal angle $\phi$ space.
For a jet to be tagged as a $b$-jet the impact parameter significance of tracks is considered.
The scalar sum of transverse object momenta and missing transverse energy has to exceed 500~GeV.
The signal region is defined by at least four (two) jets in the same-sign dilepton (trilepton)
channel. Top quark and Drell-Yan production constitute the dominant backgrounds which are determined
by means of data. Top quark plus boson and diboson production background is determined by simulation.
Exclusion limits at 95\% C.L. on the $b'\bar{b'}\rightarrow tW^-\bar{t}W^+$
production cross section are set and translated into an exclusion limit of $b'$ masses below
600~GeV.
 
\noindent
Further searches in lepton plus jet production~\cite{atlas-hepex1112.4828}~\cite{atlas_cds1389822}~\cite{atlas-hepex1204.1265}~\cite{atlas-hepex1203.5420} have been discussed.

\section{Searches in jet production} \label{jetProd}

The ATLAS search~\cite{atlas-hepex1112.5755} for heavy vector-like quarks $Q$ makes use of 
${\cal{L}}=1.04$~fb$^{-1}$. The analysis is sensitive to the charged current via
the process $pp\rightarrow Qq\rightarrow Wqq'$ and the neutral current via the process
$pp\rightarrow Qq\rightarrow Zqq'$ with leptonic decay of the vector boson.
If vector-like quarks exist they are expected to couple in general only to the third generation 
sizably. A coupling $\tilde{\kappa}_{qQ}$ is introduced to describe the model dependence
of the $qVQ$ vertex, with $V$ being one of the vector bosons $W$ or $Z$.
Events with at least two jets and a leptonically decaying vector boson are selected.
The dominating background is vector boson plus jet production, followed by 
top and diboson production which are determined from simulation.
Multijet background is estimated from data.
Jets are determined by means of the anti-$k_T$ algorithm with distance measure $R=0.4$.
95\% C.L. exclusion limits on the production cross section times branching ratio into a vector boson
plus jet have been set. 
Assuming the coupling strengths $\tilde{\kappa}^2_{uD}=1$ and $\tilde{\kappa}^2_{uU}=1$
and  the branching ratio $BR(Q\rightarrow W/Z + \mbox{jet})=100\%$,
heavy quark masses $m_Q$ below 900~GeV in the charged current and below 760~GeV 
in the neutral current can be excluded.

\noindent
Further new searches~\cite{cms-exo-11-017}~\cite{cms-exo-10-009} in jet production have been presented
as well as the long-lived particle searches~\cite{cms-exo-11-022}~\cite{atl-com-phys-2011-956}.

\section{Conclusions}
Various CMS and ATLAS searches for new phenomena
have been presented here. Complete tables of exclusion limits for all existing
analysis channels can be found in ref.~\cite{NP_MoriondQCD2012}, pp31.

\end{document}